\pdfoutput=1
\documentclass[a4paper,11pt]{article}
\usepackage{jheppub}

\usepackage{amssymb,amsmath}

\usepackage[normalem]{ulem}
\usepackage{slashed}
\usepackage{graphicx}
\usepackage{xcolor}

\usepackage{csquotes} 
\usepackage{comment}
\usepackage{mathrsfs}
\usepackage{float}

\newcommand{\bs}{\mathbf}
\newcommand{\fr}{\frac}
\renewcommand{\(}{\left(}
\renewcommand{\)}{\right)}

\newcommand{\der}{\partial}

\newcommand{\vevs}[1]{\langle #1 \rangle}

\numberwithin{equation}{section}

\preprint{
\begin{minipage}{5cm}
\small
\flushright
KYUSHU-HET-287 
\end{minipage}}

\title{Stringy constraints on primordial electromagnetic fields in axion inflation}

\author{Hajime Otsuka$^{1}$ and} 
\author{Ryo Yokokura$^{2}$} 
\affiliation{
$^1$Department of Physics, Kyushu University, 744 Motooka, Nishi-ku, Fukuoka 819-0395, Japan}
\affiliation{
$^2$Depertment of Physics \& Research and Education Center for Natural Sciences, Keio University, Hiyoshi 4-1-1, Yokohama, Kanagawa 223-8521, Japan}
\emailAdd{otsuka.hajime@phys.kyushu-u.ac.jp}
\emailAdd{ryokokur@keio.jp}

\abstract{
We study primordial electromagnetic fields in effective actions of string theory. 
In contrast to a conventional scenario of producing primordial electromagnetic fields induced by the axion inflation, 
we deal with the Dirac-Born-Infeld action as a non-linear generation of Maxwell theory. 
It turns out that the intensity of generated electromagnetic fields is bounded from above by the string scale which can also be rewritten 
in terms of supersymmetry breaking scale in the context of type IIB Large Volume Scenario. 
The instability parameter 
$\xi$ 
is constrained by the tadpole cancellation condition of D3-branes and a realization of hierarchy between the string scale and the Hubble scale of inflation. 
Hence, the magnetogenesis can be realized in the limited corner of the string landscape due to the ${\cal O}(1)$ value of the coefficient of Chern-Simons coupling.
}
\makeatletter
\gdef\@fpheader{}
\makeatother

\begin{document}

\maketitle
\setcounter{footnote}{0}%
\renewcommand{\thefootnote}{$*$\arabic{footnote}}

\section{Introduction}
Inflationary models in the early Universe
have been extensively studied in past decades.
Among others, axion inflation~\cite{Freese:1990rb,Adams:1992bn,McAllister:2008hb}, i.e., 
inflation with an axion inflaton has several features 
from the theoretical and observational viewpoints.
One is that the flatness of the inflaton potential can be originated from a classical shift symmetry of the axion,
because the axion can be understood as a Nambu-Goldstone boson.
The shift symmetry can be explicitly broken by some effects 
such as non-perturbative effects.
It is possible to generate a potential which is consistent 
with observational data~\cite{Planck:2018jri}.

Another feature is that the axion inflation models 
can generate $U(1)$ gauge fields
tachyonically.
There can be topological couplings between the axions and $U(1)$ gauge fields.
The couplings can be originated from a chiral anomaly
of the shift symmetry of the axions.
In the presence of the non-zero axion velocity, the topological couplings become first order spatial derivatives, 
and give tachyonic modes in infrared region~\cite{Carroll:1989vb,Anber:2006xt}.
The generated electromagnetic fields can be stable 
because it has a topological charge called the magnetic helicity (see e.g., Ref.~\cite{Davidson2001}).
This mechanism is called the chiral instability,
which has been discussed and 
generalized in various fields in 
modern physics~\cite{Akamatsu:2013pjd,Ohnishi:2014uea,Yamamoto:2023uzq}  
(see also a recent review~\cite{Kamada:2022nyt}).
If we consider the chiral instability in 
the axion inflation models,
the maximum of the energy density 
for the electromagnetic fields 
can be greater than the Hubble scale,
and this mechanism has been applied to 
explain the primordial magnetic fields~\cite{Turner:1987bw,Garretson:1992vt,Barnaby:2011qe,Fujita:2015iga,Anber:2015yca,Adshead:2016iae} or 
baryogenesis \cite{Giovannini:1997eg,Bamba:2006km,Anber:2015yca,Fujita:2016igl,Kamada:2016eeb,Kamada:2016cnb,Cado:2016kdp,Jimenez:2017cdr} 
(see, e.g., Refs. \cite{Durrer:2013pga,Subramanian:2015lua} for reviews of primordial magnetic fields).

Furthermore, the axion can be ubiquitous in effective 
theories derived from string theory.
String theory is one of the plausible candidates for 
quantum gravity, and it may give us an ultra-violet completion 
of inflationary models.
In string theory, there are various anti-symmetric tensor fields, and these zero-modes may be regarded as axions after the compactification of the extra dimensions.
The low-energy effective actions of the axions and $U(1)$ 
gauge fields can be obtained in the weak coupling regime (see, e.g., Refs.~\cite{Blumenhagen:2006ci,Ibanez:2012zz}).

If we construct an axion inflation model 
from string theory, we have several constraints on 
the hierarchy of energy scales such as
\begin{equation}
M_{\rm axion} < H 
< V_{\rm max}^{1/4}
< M_{\rm comp.} < M_s  < 
M_{\rm Pl}.
\label{eq:hierarchy}    
\end{equation}
Here, 
$M_{\rm axion}$, $H$, 
$V_{\rm max}^{1/4}$,
$M_{\rm comp.}$, $M_s$, and $M_{\rm Pl}$
denote a mass of the axion inflaton, 
a Hubble scale during the inflation, 
a maximum energy scale of the inflaton potential,
a scale of the compactification for the extra dimensions, 
string scale, and the reduced Planck mass, respectively.
This hierarchy, in particular the 
compactification scale, may constrain the energy scale 
of the produced electromagnetic fields via the 
chiral instability so that the field theoretical 
description works.
Therefore, it will be a non-trivial question whether 
the stringy axion inflation models can lead to 
the tachyonic production of the electromagnetic fields 
whose energy scale is much greater than the Hubble scale.
 
In this paper, we discuss the constraints on the production of the electromagnetic fields via the chiral instability 
in axion inflation models from the viewpoint of string theory.
We consider the inflationary models obtained from type IIB string theory,
where realistic cosmological models can be successfully obtained.
In Sec. \ref{sec:review}, we first review the generation of primordial electromagnetic fields induced by the non-zero axion velocity. 
In the conventional scenario, we deal with a model with quadratic kinetic terms of the axion and photon with the second order of the field strengths. 
We discuss the constraints on the produced electromagnetic fields from string theory
in Sec.~\ref{sec:DBI}.
From the viewpoint of string theory incorporating open strings, such a model will be regarded as a low-energy description of Dirac-Born-Infeld (DBI) action as reviewed in Sec. \ref{sec:revDBI}. In Sec. \ref{sec:bound}, we show that the scale of intensity of electromagnetic fields is bounded from above by the string scale; otherwise one has to deal with a non-linear generalization of Maxwell theory which is a regime beyond the validity of the corresponding effective field theory. 
The upper bound on the gauge field strength can be rewritten in terms of the Hubble scale during the inflation and the supersymmetry breaking scale in type IIB Large Volume Scenario \cite{Balasubramanian:2005zx,Conlon:2005ki} in Sec. \ref{sec:without}. 
Since all the energy scales in string theory depend on the string moduli, an amount of primordial electromagnetic fields is 
related to the sparticle masses constrained by the collider experiments. 
In Sec. \ref{sec:with}, we analyze the backreaction of electromagnetic fields induced by the axion inflation on the basis of the DBI action. 
So far, we have not specified concrete axion inflationary models. 
In Sec. \ref{sec:weak}, we consider models obtained from type IIB string theory, and examine whether 
the above constraint can be satisfied or not.
We put the constraints on the intensity of electromagnetic fields in the stringy models. 
Since the ratio the gauge coupling to the axion decay constant is uniquely determined in weakly-coupled string theory \cite{Choi:1985je,Banks:2003sx,Svrcek:2006yi}, the intensity of electromagnetic fields can be fixed by a coefficient of axion-photon coupling and the slow-roll parameter. 
Interestingly, the coefficient of axion-photon coupling is restricted to satisfy the cancellation condition of D-brane charges. 
Finally, Sec. \ref{sec:con} is devoted to the conclusions and discussions.


\section{Primordial electromagnetic fields in axion inflation}
\label{sec:review}

In this section, we review the generation of electromagnetic fields induced by the axion inflation following Refs. \cite{Turner:1987bw,Garretson:1992vt,Barnaby:2011qe,Fujita:2015iga,Anber:2015yca,Adshead:2016iae}, paying attention to the dynamics of a $U(1)$ gauge field coupled with the axion topologically.
In this paper, we do not specify 
whether the $U(1)$ gauge field
is the hypercharge gauge boson of the Standard Model 
or a hidden $U(1)$ gauge field.
Meanwhile, we consider both cases with and 
without massless fermions charged under 
the $U(1)$ gauge field, which necessarily 
exist in the Standard Model, but do not 
necessarily in the hidden sector.

First, we consider the generation of 
the electromagnetic fields in the absence of 
$U(1)$ charged massless fermions.
In the conventional scenario to produce the gauge field on the de Sitter background, the relevant action is 
\begin{align}
    S &= S_{\rm axion} + S_{\rm em},
\end{align}
with
\begin{align}
     S_{\rm axion}&= \int d^4 x \left\{\sqrt{-\det(g_{\mu \nu})} \biggl[ 
     - \frac{1}{2}g^{\mu \nu}\partial_\mu \phi \partial_\nu \phi - V(\phi) \biggl]\right\},
     \nonumber\\
     S_{\rm em}& = \int d^4 x \left\{
     \sqrt{-\det(g_{\mu \nu})}
     \left[ - \fr{1}{4g^2}
     F_{\mu\nu} F^{\mu \nu}\right]
     +
\frac{L \phi }{16\pi^2 f} 
F_{\mu \nu} \tilde{F}^{\mu \nu}
\right\}
.
\end{align}
Here, $\phi$ is an axion field with the decay constant $f$, 
$F_{\mu \nu } = \partial_\mu A_\nu - \partial_\nu A_\mu $
is the field strength of the $U(1)$
gauge field $A_\mu = (- A^0 , \mathbf{A})$
with the gauge coupling $g$,
$\tilde{F}_{\mu\nu } = \frac{1}{2}\epsilon^{\mu\nu\rho \sigma} F_{\rho \sigma} $ 
with $\epsilon^{0123} = +1 $
is the Hodge dual of the 
field strength,
$L$ is an integer parameter
and assumed to be $L > 0 $ without loss of generality,
and 
$V(\phi)$ is an axion inflaton potential.
Note that $F_{\mu \nu} \tilde{F}^{\mu \nu}
= -4 \mathbf{E} \cdot \mathbf{B}$.

The background metric $g_{\mu \nu}$ ($\mu,\nu = 0,1,2,3$) is chosen as the Friedmann-Lemaître-Robertson-Walker background with vanishing curvature:
\begin{align}
    ds^2 = -dt^2 + a^2(t) d\mathbf{x}^2 = -a^2(\eta) (d\eta^2 - d{\bf x}^2),
\end{align}
where $a$ is the scale factor and $\eta$ is the conformal time. 
Since the inflaton develops a time-dependent vacuum expectation value, the conformal invariance of the gauge field is lost; thereby the gauge field quanta will be produced during the inflation. 
It will be convenient to 
rewrite the Maxwell term with 
the canonically normalized 
kinetic term,
\begin{align}
   S = \int d^4 x \left\{ \sqrt{-\det(g_{\mu \nu})} \biggl[ -\frac{1}{2}g^{\mu \nu}\partial_\mu \phi \partial_\nu \phi - V(\phi) - \frac{1}{4}g^{\mu \rho}g^{\nu \sigma}\hat{F}_{\mu \nu}\hat{F}_{\rho \sigma} \biggl]   + \frac{\alpha L}{4 \pi f} 
   \phi
   \hat{F}_{\mu \nu} \hat{\tilde{F}}^{\mu \nu}\right\}.
\end{align}
Here, $\hat{A}_\mu$ is  a $U(1)$ gauge field 
with the canonically normalized kinetic term,
whose index is raised and/or lowered by the metric $g^{\mu \nu}$.
The parameter $\alpha = \frac{g^2}{4\pi}$
is the fine structure constant.
When we redefine the gauge field in the comoving coordinate:
\begin{align}
    \hat{A}_\mu  = (- A^0, {\bf A}) =: A_\mu,\qquad
    \hat{A}^\mu  = \left( \frac{A^0}{a^2}, \frac{{\bf A}}{a^2} \right) =: \frac{A^\mu}{a^2},
\end{align}
the action is simplified as
\begin{align}
    S = \int d^4 x \left\{\sqrt{-\det(g_{\mu \nu})} \biggl[  - \frac{1}{2}g^{\mu \nu}\partial_\mu \phi \partial_\nu \phi - V(\phi)\biggl] - \frac{1}{4}F_{\mu \nu}F^{\mu \nu}   + \frac{\alpha L}{4 \pi f} \phi F_{\mu \nu} \tilde{F}^{\mu \nu}\right\},
\end{align}
where it is notable that the index of comoving (physical) field $A$ ($\hat{A}$) is raised and/or lowered by the metric $\eta^{\mu\nu}$ ($g^{\mu \nu}$).  

The equation of motion for the $U(1)$ gauge field is 
\begin{align}
    \partial_{\mu} \left( F^{\mu \nu} + \partial^\nu A^\mu \right) = \partial_\mu \left(\frac{\alpha L}{\pi f} \phi \tilde{F}^{\mu\nu}\right),    
\end{align}
where we adopt the Feynman gauge. 
When the axion motion is negligible, the gauge field is quantized in the usual way:
\begin{align}
    A_{\mu} (x) = \int \frac{d^3 k}{(2\pi)^{3/2}} \sum_{\sigma= \pm, {\rm L}, {\rm S}}
    \biggl[ \epsilon_{\mu}^{(\sigma)} a_{\bf k}^{(\sigma)} A_\sigma (\eta, {\bf k}) e^{i {\bf k}\cdot {\bf x}} + {\rm h.c.} \biggl],
\end{align}
where $\epsilon_{\mu}^{(\sigma)}$ is the polarization vector, and the operator $\{a_{\bf k}^{(\pm)}, a_{\bf k}^{(\pm)\dagger} \}$ satisfies the commutation relation $[a_{\bf k}^{(\pm)}, a_{\bf q}^{(\pm)\dagger}] = \delta^3 ({\bf k} - {\bf q})$. 
Then, the equation of motion for $A_{\pm}$ reduces to 
\begin{align}
    \partial_\eta^2
    A_{\pm} (\eta, {\bf k}) + k (k \pm 2 
    \lambda 
    \xi a H) A_{\pm} (\eta, {\bf k}) = 0
\end{align}
with $\lambda = {\rm sign} (\dot{\phi}) $
and 
\begin{align}
    \xi = \frac{ L \alpha \lambda \dot{\phi}}{2\pi f H},
\end{align} 
where $\dot{\phi}=\fr{d\phi}{d t}$ and $H=a^{-1}\fr{d a}{d t}$. 
Thus, one of the helicities of the gauge field will be a tachyonic mode for $k< 2aH\xi $. 
If we take $\lambda >0 $, the tachyonic mode is $A_-$. 
As we will see below, the parameter $\xi $ characterizes the intensity of the electromagnetic fields.
This parameter is called the instability parameter (see e.g., Ref.~\cite{Jimenez:2017cdr}),
which plays an important role in discussing the 
generation of the electromagnetic fields by the instability as we will see below.
Note that we assume that the instability parameter 
can be approximately a constant, which will be justified 
during and at the end of the inflation.

By imposing the plane wave for $A_{\pm} \propto e^{ -i \omega_k \eta }$
with
 $\omega_k^2 = k (k \pm 2 \lambda a H)$
in an enough inside the horizon ($-k\eta \gg 1$), the gauge field is obtained as
\begin{align}
    A_{-\lambda}(\eta, {\bf k}) = \frac{e^{\pi \xi/2}}{\sqrt{2k}} W_{-i\lambda \xi, 1/2} (2ik\eta).
\end{align}
As a result, the physical electric and magnetic fields 
as well as the correlation length of the magnetic fields
$\lambda_B$ are estimated as~\cite{Jimenez:2017cdr}
\begin{align}
    \langle \hat{E}^2\rangle &=
     \frac{e^{2\pi \xi}}{|\xi|^3} H^4 \left( \frac{|\xi|^3}{4\pi^2}e^{-\pi \xi} \int_0^{\kappa_{\rm UV}} d\kappa\, \kappa^3 \frac{\partial}{\partial \kappa} | W_{-i\lambda \xi, 1/2} (-2i\kappa)|^2\right)
    \nonumber\\
    &\simeq 2.6\times 10^{-4}  \frac{e^{2\pi \xi}}{|\xi|^3} H^4,
    \nonumber\\
    \langle \hat{B}^2\rangle &= \frac{e^{2\pi \xi}}{|\xi|^5} H^4 \left( \frac{|\xi|^5}{4\pi^2}e^{-\pi \xi} \int_0^{\kappa_{\rm UV}} d\kappa\, \kappa^3 \frac{\partial}{\partial \kappa} | W_{-i\lambda \xi, 1/2} (-2i\kappa)|^2\right)
    \nonumber\\
    &\simeq 3.0\times 10^{-4} \frac{e^{2\pi \xi}}{|\xi|^5} H^4,
    \nonumber
    \\
    \langle \mathbf{E}
    \cdot \mathbf{B}
    \rangle 
    & = 
        \lambda \frac{e^{2\pi \xi}}{|\xi|^4} H^4 \left(  - \frac{|\xi|^4}{8\pi^2}e^{-\pi \xi} \int_0^{\kappa_{\rm UV}} d\kappa\, \kappa^3 \frac{\partial}{\partial \kappa} | W_{-i\lambda \xi, 1/2} (-2i\kappa)|^2\right)
    \nonumber\\
    &\simeq 2.6\times 10^{-4} \lambda \frac{e^{2\pi \xi}}{|\xi|^4} H^4,
       \nonumber
    \\
    \lambda_B
    & 
    = 
         \frac{4 \pi  e^{2\pi \xi} H^3}{|\xi|^4 \vevs{\hat{B}^2} } \left(  \frac{|\xi|^4}{8\pi^2}e^{-\pi \xi} \int_0^{\kappa_{\rm UV}} d\kappa\, \kappa^2
         | W_{-i\lambda \xi, 1/2} (-2i\kappa)|^2\right)
    \nonumber\\
    &\simeq 6.0 \times 10^{-1} \frac{2 \pi}{H},
\end{align}
where we take $\kappa = -k \eta = k/(aH)$ and $\kappa_{\rm UV} \geq 6$. 
The intensity of electromagnetic fields is
\begin{align}
    \langle \hat{F}_{\mu \nu}\hat{F}^{\mu \nu}\rangle = 2(\langle \hat{B}^2\rangle - \langle \hat{E}^2\rangle) \simeq 6\times 10^{-4} e^{2\pi \xi}H^4 \left( \frac{1}{|\xi|^5} - \frac{1}{|\xi|^3}  \right),
    \label{eq:intensity}
\end{align}
which is highly dependent on $\xi$. 
When $|\xi| \gtrsim {\cal O}(1)$, one of the gauge field is amplified due to the tachyonic instability, and the large production of electromagnetic fields leaves a measurable imprint on cosmological observables~\cite{Anber:2006xt,Jimenez:2017cdr},
while we should be careful about the 
screening by 
charged particles~\cite{Domcke:2018eki,Sobol:2019xls,Fujita:2022fwc}
or backreactions~\cite{Sobol:2019xls,vonEckardstein:2023gwk}.

We should remark that the scale of the intensity or the energy scale of the electromagnetic fields can exceed the Hubble scale.
However, the energy of the electromagnetic fields 
cannot exceed the maximum value 
of the inflaton potential $V_{\rm max}^{1/4}$.
This constraint further implies that it cannot be greater than 
$M_{\rm comp.}$
and 
$M_s$
to preserve the hierarchy~\eqref{eq:hierarchy}
if we consider four-dimensional (4D) models using string theory.

Next, we include massless fermions charged under the $U(1)$ gauge field~\cite{Domcke:2018eki}.
Since the electromagnetic fields produce the massless fermions by the Schwinger mechanism,
and the generation of the electromagnetic fields is non-trivially changed.
For simplicity, we consider 
a Dirac fermion $\psi$ described by the Lagrangian,
\begin{equation}
 {\cal L}_{\rm f} = i \bar{\psi}  \gamma^a e_a{}^\mu
\(\der_\mu + \fr{1}{4} \omega_\mu{}^{ab} \Sigma_{ab} + i 
A_\mu \) \psi,
\end{equation}
where $\gamma_a $, ($a = 0,1,2,3$) is the Gamma matrix,
with $\{\gamma_a , \gamma_b\} = 
 - 2\eta_{ab}$,
$\Sigma_{ab} = \fr{1}{2}[\gamma_a,\gamma_b ]$
is the generator of the Lorentz group acting on the Dirac fermion, 
$e_a{}^\mu$ is the vielbein satisfying 
$\eta_{ab} = e_a{}^\mu e_b{}^\nu g_{\mu\nu}$,
and $\omega_\mu{}^{ab}$ is the spin connection made of the vielbein.
Since the Dirac fermion is massless, there is classically 
the axial symmetry denoted as $U(1)_A$
described by the conserved current
$j_{A}^\mu = \bar{\psi}  \gamma^\mu \gamma_5 \psi$, 
but it is violated by the axial anomaly,
$ \der_\mu \vevs{j_A^\mu} =  \fr{1}{2\pi^2} \bs{E} \cdot \bs{B}$.
This shows that the parallel electric and magnetic fields 
can be a source of the fermion.

While it is in general difficult to discuss the fermion production 
exactly,
it has been shown in Ref.~\cite{Domcke:2018eki} that 
we can approximately compute the production under the assumption 
that the electromagnetic fields are constant.
This can be justified because the fermion production is much faster than 
the tachyonic production of the electromagnetic fields.
Under the assumption, we can show that 
the production of the electric current 
$j^\mu = \bar{\psi} \gamma^\mu \psi$
is 
\begin{equation}
 \fr{g}{a^3} \vevs{j^3} 
\simeq
 \fr{g^3}{6\pi^2}
\fr{\hat{E}  \hat{B} }{H}
\coth \(\fr{\pi \hat{B} }{ \hat{E} } \)
,
\end{equation} 
where we have chosen the direction of the electromagnetic fields 
as the $x^3$-direction,
$\hat{\bs{E}} = (0,0,\hat{E} )$
and 
$\hat{\bs{B}} = (0,0,\lambda\hat{B} )$.
The conservation law of the electromagnetic fields 
$\rho_A = \fr{1}{2} (\hat{\bs{E}}{}^2 + \hat{\bs{B}}{}^2)$
from the Maxwell-Amp\`ere law, 
$\dot{\rho}_A = - 4 H \rho_A + 2\xi H \hat{\bs{E}} \cdot \hat{\bs{B}}
- g \hat{\bs{E}}_i \cdot \langle \bs{j}_i \rangle $,
shows that the instability parameter $\xi$ is effectively modified 
as
\begin{equation}
 \xi_{\rm eff}
: = \xi - 
 \fr{g^3}{12\pi^2}
\fr{\hat{E}  }{H^2}
\coth \(\fr{\pi \hat{B} }{ \hat{E} } \).
\end{equation}

The modification of the instability parameter gives us the upper bound on the electromagnetic fields.
The upper bound
can be estimated by imposing the assumption that the time evolution of the 
electromagnetic energy is approximately zero,
$\dot\rho_A \simeq 0$~\cite{Domcke:2018eki}.
If $\xi_{\rm eff}$ is dominated by $\xi \gg 1$, 
the condition 
$\dot\rho_A = -2 H 
(\hat{\bs{E}}{}^2 + \hat{\bs{B}}{}^2) 
 + 2 \xi_{\rm eff} H \hat{\bs{E}}\cdot\hat{\bs{B}} \simeq 0$ leads to the linear relation between 
$\hat{\bs{E}}$ and $\hat{\bs{B}}$,
$\hat{\bs{E}} \simeq \xi \hat{\bs{B}} $ 
with $|\hat{\bs{E}}| > |\hat{\bs{B}}| $.
On the other hand, this linear relation 
is violated when the correction by the Schwinger mechanism cannot be neglected.
The upper bound of the electric field can be roughly estimated around this region, 
$\xi \sim  \fr{g^3}{12\pi^2}
\fr{\hat{E}  }{H^2}
\coth \(\fr{\pi \hat{B} }{ \hat{E} } \)$, 
which implies that 
\begin{equation}
\hat{E}_{\rm max} \sim \fr{12 \pi^2}{  g^3} \xi H^2,
\label{eq:Emax_wfermion}
\end{equation}
with the assumption 
$\coth \(\fr{\pi \hat{B} }{ \hat{E}} \) \sim 1$.
The upper bound on the intensity of the electromagnetic fields is 
\begin{equation}
       |\langle \hat{F}_{\mu \nu}\hat{F}^{\mu \nu}\rangle|
       = |2(\langle \hat{B}^2\rangle - \langle \hat{E}^2\rangle)| < 
       2 H^4 \(\fr{12\pi^2 \xi}{g^3}\)^2 . 
    \label{eq:intensity-massless} 
\end{equation}
Thus, the exponential growth of the electromagnetic fields by 
the instability parameter 
can be suppressed in the presence of the massless fermion.

In the following sections, 
we will discuss the possible constraints of 
the energy scale of the electromagnetic fields characterized by the instability parameter $\xi$
from the viewpoint of string theory.
One is the constraint from the DBI action, and 
another is from the 
tadpole constraints~\cite{Bena:2020xrh,Ishiguro:2023jjc}.

\section{DBI constraints on electromagnetic fields}
\label{sec:DBI}

In this section, we will show that 
the DBI action gives us some constraints from string theory.
One is 
the possible value of the coupling constant $g$,
which is determined by 
the string coupling constant and the string tension.
Another is the 
upper bound on the generated electromagnetic fields so that 
the expansion of the DBI action is under control.
We show that the upper bound will be 
related to the Planck mass
and the volume of the extra dimensions.

\subsection{Review of DBI action\label{sec:revDBI}}
First, we briefly review the DBI action as 
known as a non-linear generalization of Maxwell theory
\footnote{Note that other higher-order corrections 
for the electromagnetic fields are possible.
For example, we can take the Euler-Heisenberg 
action, which is a 1-loop effective action 
after integrating out a massive field coupled with the gauge field~\cite{Domcke:2019qmm}.
In this case, the constraint on the magnetic field means that it should be sufficiently small so that the dynamics of the massive mode can be neglected.}.
(See, Refs. \cite{Ibanez:2012zz,Blumenhagen:2013fgp}, in more details.)
In the context of string theory, the DBI action describes the dynamics of the massless open string modes involving the electromagnetic fields which couple to the bulk fields in the Neveu-Schwarz sector, i.e., 
the dilaton $\Phi$, the metric, and the Kalb-Ramond $B$-field. 
In particular, for a single D$p$-brane with $p<9$, the bosonic part in the string frame is described by
\begin{align}
    S_{\rm DBI} = -T_p \int d^{p+1} \xi e^{-\Phi(X)}\sqrt{-\det (g_{\alpha \beta}(X) + 2\pi \alpha^\prime {\cal F}_{\alpha \beta}(X))},
    \label{eq:DBI}
\end{align}
where $T_p = 2\pi l_s^{-(p+1)}$ is the tension of D$p$-brane\footnote{Here, we consider type II string theory, but we need the extra factor $1/\sqrt{2}$ in the case of type I string theory.}, $l_s = 2\pi \sqrt{\alpha^\prime}$ is the string length, and $2\pi \alpha^\prime {\cal F} = B + 2\pi \alpha^\prime F$ is the gauge invariant field strength including the $U(1)$ gauge field strength $F$. 
Here and in what follows, we denote $A,B = 0,1,...,9$ for 10-dimensional (10D) spacetime indices, $\alpha, \beta=0,1,...,p$ for the brane world-volume coordinates, and $i,j=p+1,...,9$ for the intrinsic world volume coordinates, respectively.
Since the embedding of the world volume ${\cal W}$ into 10D spacetime is specified by the function $X^M(\xi)$, the pull-back metric $g_{\alpha \beta}$ and $B_{\alpha \beta}$ are 
respectively given by
\begin{align}
    g_{\alpha \beta} = \partial_\alpha X^A \partial_\beta X^B G_{AB},\qquad
    B_{\alpha \beta} = \partial_\alpha X^A \partial_\beta X^B B_{AB}.
\end{align}

For simplicity, let us assume the $B$-field and fluctuations of D$p$-brane to zero and 
adopt the static gauge\footnote{The fluctuations of D$p$-brane, i.e., open string moduli, will be stabilized by four-form fluxes in the context of M-theory and F-theory compactifications \cite{Becker:1996gj,Sethi:1996es,Gukov:1999ya,Denef:2005mm,Honma:2017uzn}.}:
\begin{align}
    X^\alpha = \xi^\alpha,\qquad
    X^i = x^i. 
\end{align}
In this gauge, the DBI action is expanded as follows:
\begin{align}
    S_{\rm DBI} &= -T_p \int d^{p+1} \xi e^{-\Phi(X)}\sqrt{-\det (g_{\alpha \beta})} 
    \sqrt{\det (1 + 2\pi\alpha^\prime g^{\alpha \beta}{\cal F}_{\alpha \beta})}
    \nonumber\\
    &= -T_p \int d^{p+1} \xi e^{-\Phi(X)}\sqrt{-\det (g_{\alpha \beta})} 
    \biggl[ 1+ \frac{(2\pi \alpha^\prime)^2}{4}F^a_{\alpha \beta}F^{a \alpha \beta} -\frac{(2\pi \alpha^\prime)^4}{12} {\cal O}(F^4) 
    \biggl],
\end{align}
with ${\rm tr}(T^aT^b)=\delta^{ab}$.

Since we are interested in the electromagnetic fields, 
we assume that the gauge group 
has a $U(1)$ subgroup.
The $U(1)$ gauge coupling on the D$p$-brane is then extracted as
\begin{align}
    g_{{\rm D}p}^2 = \frac{e^{\langle \Phi \rangle}}{T_p(2\pi \alpha^\prime)^2} = g_s (2\pi)^{p-2} (\alpha^\prime)^{\frac{p-3}{2}},
\end{align}
where $g_s = e^{\langle \Phi \rangle}$ 
represents the string coupling. 
Thus, the Maxwell term as well as the higher-order corrections for the gauge fields are given by
\begin{align}
    S_{\rm DBI} \supset -\int d^{p+1} \xi \sqrt{-\det (g_{\alpha \beta})} 
    \frac{1}{4g_{{\rm D}p}^2}\biggl[ F_{\alpha \beta}F^{\alpha \beta} -\frac{(2\pi \alpha^\prime)^2}{3} {\cal O}(F^4) 
    \biggl].
\end{align}

\subsection{Upper bound on electromagnetic fields from DBI action}
\label{sec:bound}

Now, we are ready to discuss a constraint on the electromagnetic fields originating from the DBI action.
In the conventional scenario to generate the 4D electromagnetic fields in the early Universe, 
we often deal with the quadratic (Maxwell) term of the $U(1)$ gauge field. 
However, the non-linear generalization of the Maxwell theory is motivated in string theory. 
For concreteness, we consider 
the DBI action for a D3-brane system in type IIB string theory.
\footnote{When we start from the higher-dimensional DBI action, one can introduce the electric or magnetic field in the extra-dimensional space. 
They are typically of order unity in the unit of the string length, otherwise the tadpole cancellation conditions will be violated. 
Thus, the critical value of the 4D electromagnetic field strength (\ref{eq:Fbound}) is still controlled by the volume of extra-dimensional spaces, and the following discussion is applicable to such a system. We will come back to this point in Sec.~\ref{sec:weak}.}
The 4D electromagnetic field strength is bounded from above as:
\begin{align}
    F^2_{\rm max} \simeq \frac{1}{(2\pi \alpha^\prime)^2} \equiv (2\pi)^2 M_s^4,
    \quad 
    \text{or equivalently,}
    \quad
       \hat{F}^2_{\rm max} \simeq \frac{1}{(2\pi \alpha^\prime)^2g_{{\rm D}3}^2} = 2\pi \frac{M_s^4}{g_s} \simeq \frac{M_{\rm Pl}^4}{8\pi{\cal V}^{2}}
  ;
    \label{eq:Fbound}
\end{align}
otherwise, the DBI action in \eqref{eq:DBI} is not justified because the action becomes imaginary. 
Indeed, when the electromagnetic fields approach this critical value, 
one cannot neglect the 
Schwinger effect for strings \cite{Bachas:1992bh}. 
Note that the string scale $M_s = (l_s)^{-1} = (2\pi \sqrt{\alpha^\prime})^{-1}$ is written in terms of the 4D reduced Planck mass $M_{\rm Pl}$ and the Einstein-frame volume of extra six-dimensional space ${\cal V}$ in units of the string length:
\begin{align}
    M_s = \frac{1}{l_s} \simeq \frac{g_s^{1/4}}{\sqrt{4\pi {\cal V}}}M_{\rm Pl}.
    \label{eq:string-volume-Planck}
\end{align}
Note that this relation can be obtained 
from the reduction of the 10D Einstein-Hilbert action 
in the string frame
to the 4D one,
$S_{\rm EH, 10D} = \fr{2 \pi}{g_s^2 l_s^8} 
\int d^{10} x 
\sqrt{-g^{\rm 10D}} R^{\rm 10 D}
\to S_{\rm EH, 4D} 
 = 
 \fr{2 \pi {\cal V}_s}{g_s^2 l_s^2} 
\int d^4 x 
\sqrt{-g }R^{\rm 4D} = 
\fr{M_{\rm Pl}^2}{2} \int d^4 x 
\sqrt{-g }R^{\rm 4D}$.
Here,
$g^{\rm 10D}_{AB}$ 
and $g^{\rm 10D}$ 
are the metric in 10D and its determinant, respectively. 
The symbols 
$R^{\rm 10 D}$ and $R^{\rm 4D}$
denote the Ricci scalar 
in 10D and 4D, respectively.
${\cal V}_s = \fr{1}{l_s^6}\int d^6 x
\sqrt{g^{\rm 6D}}$ is the volume of 
the extra dimensions in the string frame, which is related to ${\cal V}$
in the Einstein frame as 
${\cal V}_s = g_s^{3/2} {\cal V} $.

From the upper bound, 
we can find two constraints.
One is the upper bound of the instability parameter $\xi$.
The other is 
the lower bound on the breaking scale of the 
supersymmetry
because
the bound is related to the size of the extra dimensions.
In the following, we estimate 
the upper bound on the parameter 
$\xi$, which characterizes the 
intensity of the electromagnetic fields.
When the dynamics of electromagnetic fields is governed by the D3-brane action with the Chern-Simons term, the relevant action is
\begin{align}
    S &= S_{\rm axion} + S_{\rm em},
\end{align}
with
\begin{align}
     S_{\rm axion}&= \int d^4 x \left\{\sqrt{-\det(g_{\mu \nu})} \biggl[ - \frac{1}{2}g^{\mu \nu}\partial_\mu \phi \partial_\nu \phi - V(\phi) \biggl]\right\},
     \nonumber\\
     S_{\rm em}& = \int d^4 x \left\{-T_3e^{-\Phi(x)}\sqrt{-\det(g_{\mu \nu})} \Biggl[\sqrt{1+ \frac{(2\pi \alpha^\prime)^2}{2}F_{\mu \nu}F^{\mu \nu}  -  \frac{(2\pi \alpha^\prime)^4}{16}(F_{\mu \nu}\tilde{F}^{\mu \nu})^2}\, \Biggl] \right.
     \nonumber\\
     &\hspace{2cm} +\left.\frac{\phi }{16\pi^2 f} F_{\mu \nu} \tilde{F}^{\mu \nu}\right\},
\label{eq:D3DBI}
\end{align}
where the axion field $\phi $, its the decay constant $f$,
and the axion-photon coupling 
can be 
derived from the kinetic terms 
of Ramond-Ramond (RR) fields,
and the Chern-Simons coupling
on the D3-brane,
which we will discuss in Sec.~\ref{subsec:independent}. 
$V(\phi)$ is an axion inflaton potential.
Note that we have $L =1$ for the D3-brane effective theories, because D3-branes are localized in the extra dimensions.
After canonically normalizing the gauge fields, the relevant action of the gauge fields is 
\begin{align}
    S_{\rm DBI} 
    &= -T_3 \int d^4 \xi \sqrt{-\det (g_{\alpha \beta})} \sqrt{1+ \frac{(2\pi \alpha^\prime)^2}{2}F_{\alpha \beta}F^{\alpha \beta}  -  \frac{(2\pi \alpha^\prime)^4}{16}(F_{\alpha \beta}\tilde{F}^{\alpha \beta})^2} 
    \nonumber\\
    &= -\int d^{4} \xi e^{-\Phi(X)}\sqrt{-\det (g_{\alpha \beta})} 
    \frac{1}{4}\biggl[ \hat{F}_{\alpha \beta}\hat{F}^{\alpha \beta} -\frac{(2\pi \alpha^\prime)^2g_{{\rm D}3}^2}{3} {\cal O}(\hat{F}^4)
    \biggl],
\end{align}
with $\alpha, \beta = 0,1,2,3$. 
We should remark that the expansion of the DBI action up to the quadratic order of the field strength is meaningful as long as 
$\hat{F}^2 \ll  M_{\rm Pl}^4/ (8\pi {\cal V}^2)$,
otherwise the higher-order term cannot be neglected.
In the following subsections, we first discuss the generation of $U(1)$ gauge field 
incorporating the DBI constraint (\ref{eq:Fbound}) in Sec. \ref{sec:without}. 
In Sec. \ref{sec:with}, we next examine the backreaction from the DBI action. 
\subsection{Analysis up to quadratic order of electromagnetic fields}
\label{sec:without}

After expanding the DBI action (\ref{eq:D3DBI}) up to a quadratic order of $F$, 
we can read off the axion employed in the conventional scenario 
to produce the gauge field on the de Sitter background:
\begin{align}
   S_{\rm quad} = \int d^4 x \left\{\sqrt{-\det(g_{\mu \nu})} \biggl[ -\frac{1}{2}g^{\mu \nu}\partial_\mu \phi \partial_\nu \phi - V(\phi) - \frac{1}{4}g^{\mu \rho}g^{\nu \sigma}\hat{F}_{\mu \nu}\hat{F}_{\rho \sigma} \biggl]   + \frac{\alpha_{{\rm D3}} }{4 \pi f} 
   \phi
   \hat{F}_{\mu \nu} \hat{\tilde{F}}^{\mu \nu}\right\},
\end{align}
where $\hat{A}_\mu$ is  a $U(1)$ gauge field 
with the canonically normalized kinetic term, and the index is raised and/or lowered by the metric $g^{\mu \nu}$.
The parameter $\alpha_{\rm D3}$ is the fine structure constant 
given by 
$\alpha_{{\rm D3}} = \frac{g^2_{\rm D3}}{4\pi}$.

The effective action of string theory will be justified in the small string coupling and the large volume expansions, as utilized in the Large Volume Scenario \cite{Balasubramanian:2005zx,Conlon:2005ki}. 
Since the large volume of the extra-dimensional space significantly reduces the critical value 
of the 4D electromagnetic fields, it will be quite important to incorporate this DBI bound in the 
context of the 
tachyonic production of the 
electromagnetic fields
in the early Universe. 
In particular, we focus on the generation of 4D electromagnetic fields in generic axion inflation models.

\subsubsection{Without massless fermion}

Let us examine whether the produced electromagnetic fields in the early Universe satisfy the DBI bound (\ref{eq:Fbound}). 
By utilizing Eq. (\ref{eq:intensity}), 
we arrive at the condition 
such that 
the expansion of the DBI action can be controlled against the higher-order corrections, 
\begin{align}
    \frac{e^{\pi \xi}}{|\xi|^{3/2}} &\ll 
    324 \left(\frac{0.1}{g_s}\right)^{1/2} \left(\frac{M_s}{H} \right)^2 \simeq 
    8.1\times 10^5\left(\frac{10^{7}}{{\cal V}}\right) \left( \frac{10^{-6}\,M_{\rm Pl}}{H}\right)^2,
\label{eq:xi}
\end{align}
indicating that the large volume of the extra-dimensional space strongly constrains the amount of gauge fields. 
Here, the string scale is chosen to be close to the grand unification scale 
while respecting the hierarchy in 
Eq.~\eqref{eq:hierarchy},
$10^{14} \leq M_s \leq 10^{16}$\,GeV, taking into account the uncertainty of high-scale threshold corrections. 

The volume range is now within the range $10^3 \leq {\cal V} \leq 10^7$ for $g_s \simeq 0.1$
due to the relation in \eqref{eq:string-volume-Planck}. 
Such a large volume is preferred in the string compactification due to 
the fact that the perturbative string effective field theory is controlled by the volume expansion. 
As a non-trivial consequence, 
it is possible to constrain 
the lower bound on the supersymmetry breaking scale.
Here, we consider the Large Volume Scenario, where all moduli can be stabilized 
and 
de Sitter uplifting can be realized.
Indeed, in the Large Volume Scenario, the volume modulus is stabilized at the 
supersymmetry breaking minimum with exponentially large volume. 
In particular, the gravitino mass is given by~\cite{Conlon:2005ki}\footnote{Here, the vacuum expectation value of K\"ahler potential for the complex structure moduli is chosen as $e^{\fr{K_{\rm cs}}{2M_{\rm Pl}^2}}=1$.}
\begin{align}
m_{3/2}=\left( \frac{g_s^2}{2\sqrt{2\pi}}\right)\frac{|W_0|}{\mathcal{V}}M_{\rm Pl},
\label{eq:const1}
\end{align}
where $W_0$ is the (dimensionless) vacuum expectation value of the flux-induced superpotential $W$. 
Thus, the constraint (\ref{eq:const1}) for electromagnetic fields in the early Universe is also rewritten in terms of the supersymmetry breaking scale rather than the volume:
\begin{align}
     \frac{e^{\pi \xi}}{|\xi|^{3/2}} &\ll 8.1\times 10^5 \left( \frac{50}{|W_0|}\right)\left( \frac{0.1}{g_s}\right)^2\left( \frac{m_{3/2}}{2.4\times 10^{10}\,{\rm GeV}}\right) \left( \frac{10^{-6}\,M_{\rm Pl}}{H}\right)^2.    
     \label{eq:Upperbound}
\end{align}
Note that the magnitude of the flux superpotential is statistically favored at $|W_0| = {\cal O}(10)$ \cite{Cicoli:2013swa}. Of course, it has an option to consider small $|W_0|$ required in the KKLT construction, but it will be a limited corner of the string landscape.
\footnote{See for a realization of the small flux superpotential by instanton effects, e.g., Refs.~\cite{Demirtas:2019sip,Demirtas:2020ffz,Alvarez-Garcia:2020pxd,Honma:2021klo}.} 

Let us apply the condition \eqref{eq:Upperbound} to the production of the gauge field via the chiral instability.
To produce a large amount of gauge field quanta during the axion inflation, for instance, $\frac{e^{\pi \xi}}{|\xi|^{3/2}} \simeq 6\times 10^5$ for $\xi \sim 5$ used in a recent analysis of Ref. \cite{vonEckardstein:2023gwk}, 
the above inequality indicates that the typical supersymmetry breaking scale (i.e., the gravitino mass) is at least larger than ${\cal O}(10^{10})$ GeV scale; 
otherwise one has to incorporate non-linear effects.

Since the lower bound on the supersymmetry breaking scale can be estimated, 
one can further discuss the typical mass scale of the sparticles.
In string models, the scale of soft supersymmetry breaking terms depends on the location and type of Standard Model D-brane configuration. 
For illustrative purposes, let us focus on the sequestered model \cite{Higaki:2012ar,Aparicio:2014wxa} where the Standard Model sector will be realized at a certain singularity of the underlying Calabi-Yau orientifold. The explicit model building on concrete Calabi-Yau orientifold with Minimal Supersymmetric Standard Model (MSSM)-like D3-branes at singularities was performed in type IIB flux compactifications with O3/O7-planes \cite{Cicoli:2012vw,Cicoli:2013mpa,Cicoli:2013cha,Cicoli:2017shd,Cicoli:2021dhg}. The sparticle spectrum depends on the mechanism of uplifting the anti de-Sitter minimum in the Large Volume Scenario to a de Sitter vacuum as analyzed in Ref. \cite{Aparicio:2014wxa}. When we focus on a specific uplifting scenario supported by E$(-1)$-instantons \cite{Cicoli:2012fh}, one can realize both a typical MSSM-like and split supersymmetry spectra: 
\begin{enumerate}
    \item Typical MSSM scenario

    In the so-called ultra-local scenario, the gaugino mass $M_{1/2}$, soft scalar masses $m_0$ are the same order:
    \begin{align}
        M_{1/2} \sim m_0 \sim m_{3/2}\varepsilon \ll m_{3/2},
    \end{align}
    with $\varepsilon \sim {\cal V}^{-1}$. 
    Thus, even if the gravitino mass is of ${\cal O}(10^{10})$ GeV, one can realize ${\cal O}(1)$ TeV soft masses due to the large volume suppression ${\cal V}\simeq 10^7$. 
    However, this scenario will be in tension with bounds on charge and color breaking vacua in the MSSM \cite{Cicoli:2023gbd}. 

        \item Split supersymmetry scenario

    In the so-called local model, the gaugino and soft scalar masses are also suppressed by the Calabi-Yau volume:
    \begin{align}
        M_{1/2}&\sim m_{3/2} \varepsilon,
        \nonumber\\
        m_0 &\sim m_{3/2}\sqrt{\varepsilon}. 
    \end{align}
    The large volume ${\cal V}\simeq 10^7$ and  ${\cal O}(10^{10})$ GeV gravitino mass lead to the spectra of split supersymmetry: $M_{1/2}\sim {\cal O}(1)\,\text{TeV}, m_0 \sim {\cal O}(100)\,\text{TeV}$. 
\end{enumerate}
In both scenarios, Eq. (\ref{eq:Upperbound}) indicates that the supersymmetry breaking scale, more specifically the sparticle masses constrained by the collider experiments put a bound on the amount of electric and magnetic fields.

We can further apply the DBI constraint 
to the explanation of the lower bound on the intergalactic magnetic field (IGMF) 
$B_0 \gtrsim  10^{-16} {\rm G}$ with the coherence length $\lambda_B > 10\,{\rm kpc} $ \cite{Finke:2015ona,Fermi-LAT:2018jdy}.
If we assume that the origin of the IGMF is the magnetic field generated by the axion inflation, we can calculate the 
strength and the coherence length~\cite{Jimenez:2017cdr}
\begin{align}
    \hat{B} &\simeq 2.5 \times 10^{-19}{\rm G} \left(\fr{e^{2\pi \xi}}{\xi^4}\right)^{1/3} \left(\fr{H_{\rm inf}}{10^{13}\,{\rm GeV}}\right)^{1/2},
    \nonumber\\
    \lambda_B &\simeq 0.28\,{\rm pc}\left(\fr{B}{10^{-14}\,{\rm G}}\right).
\end{align}
To have the magnetic field
and the coherence length 
consistent with the observations, the instability parameter will be 
$ 4 \lesssim \xi $ for $H_{\rm inf}=10^{13}\,{\rm GeV}$.
These values imply that it will be possible to explain the IGMF
by string theory
with the region $4 \lesssim \xi \lesssim 5$
taking the constraints by the DBI action into account,
but constrained to a limited corner of the string landscape.

 \subsubsection{With massless fermion}

 Next, we consider the upper bound by the DBI action in the presence of the massless fermion.
The upper bound in \eqref{eq:Fbound} 
with the intensity of the electromagnetic fields in \eqref{eq:intensity-massless}
shows that the intensity should be sufficiently small,
\begin{equation}
2(12\pi^2 \xi)^2
    \ll (2 \pi)^4 g_s^2 \fr{M_s^4}{H^4} 
    \simeq 15.6 
    \left( 
    \frac{g_s}{0.1} 
 \right)^2 \left( 
    \fr{M_s}{H} 
 \right)^4 ,
        \label{eq:massless-upper-bound} 
\end{equation}
where we have used $g^2 = g_{\rm D3}^2 = 2 \pi g_s$.
The constraints on $\xi$ in this case 
is not so strict if we take the mass hierarchy into account, e.g., 
$\xi \ll 2.4 \times 10^4$ for $M_s \simeq 10^3 H$.

\subsection{Analysis beyond quadratic order of electromagnetic fields}
\label{sec:with}

So far, we have analyzed the $U(1)$ gauge field governed by the DBI action up to a quadratic order. 
It is however non-trivial how higher order terms affect our discussion.
To see this, we deal with the DBI action by fluctuating the gauge fields at a given time:
\begin{align}
    F_{\mu \nu} \rightarrow \langle F_{\mu \nu} \rangle + \delta F_{\mu \nu},
\label{eq:fluctuation}
\end{align}
where $\langle F_{\mu \nu} \rangle$ is the background gauge field strength obtained in the previous section. 
For simplicity, in the following analysis, we set the background gauge field to time-independent; the gauge field quanta is evaluated at a given time. 

To analytically evaluate the backreaction, we start from the DBI action in the limit of $\alpha^\prime \ll 1$:
\begin{align}
    S_{\rm DBI}& = -T_3e^{-\langle\Phi\rangle}\int d^4 x \sqrt{-\det(g_{\mu \nu})} \sqrt{1+ \frac{(2\pi \alpha^\prime)^2}{2}F_{\mu \nu}F^{\mu \nu}}.
\end{align}
By fluctuating the gauge field (\ref{eq:fluctuation}), one can derive the action of 
$\delta F_{\mu \nu}$ up to a quadratic order:
\begin{align}
    S_{\rm DBI}& = -T_3e^{-\langle\Phi\rangle}\int d^4 x \sqrt{-\det(g_{\mu \nu})} \sqrt{1+ \frac{(2\pi \alpha^\prime)^2}{2}F_{\mu \nu}F^{\mu \nu}}
    \nonumber\\
    &= -T_3e^{-\langle\Phi\rangle}\int d^4 x \sqrt{-\det(g_{\mu \nu})} \sqrt{1+ \frac{(2\pi \alpha^\prime)^2}{2}\left( \langle F_{\mu \nu} \rangle + \delta F_{\mu \nu}\right)\left( \langle F^{\mu \nu} \rangle + \delta F^{\mu \nu}\right)}
    \nonumber\\
    &= -\int d^4 x \sqrt{-\det(g_{\mu \nu})} 
\frac{1}{4g_{{\rm D}3}^2\sqrt{1+ \frac{(2\pi \alpha^\prime)^2}{2} \langle F_{\mu \nu} F^{\mu \nu} \rangle}} \delta F_{\mu \nu}\delta F^{\mu \nu} + {\cal O}(\delta F^4) .
\end{align}
Thus, we arrive at the action of the fluctuation $\delta F$:
\begin{align}
    S_{\rm em}^{(\delta F)} = -\int d^4 x \sqrt{-\det(g_{\mu \nu})} 
\frac{1}{4g_{{\rm D}3}^2\sqrt{1+ \frac{(2\pi \alpha^\prime)^2}{2} \langle F_{\mu \nu} F^{\mu \nu} \rangle}} \delta F_{\mu \nu}\delta F^{\mu \nu} + \frac{\langle \phi\rangle }{4f} \delta F_{\mu \nu} \delta \tilde{F}^{\mu \nu},
\end{align}
where the axion $\phi$ is assumed to be the background value. (See for the backreaction of the axion field, e.g., Ref.~\cite{Domcke:2020zez}.)
The equation of motion for the fluctuation $\delta F$ is the same as its background, except for the gauge coupling. The corrected gauge coupling is
\begin{align}
   G_{\rm D3}^2 &:= g_{\rm D3}^2\sqrt{1+ \frac{(2\pi \alpha^\prime)^2}{2} \langle F_{\mu \nu} F^{\mu \nu} \rangle}
   = g_{\rm D3}^2\sqrt{1+ (2\pi \alpha^\prime)^2(\langle \hat{B}^2\rangle - \langle \hat{E}^2\rangle)}
   \nonumber\\
   &\simeq  g_{\rm D3}^2\sqrt{1+ \frac{3\times 10^{-4}}{(2\pi)^2} e^{2\pi \xi}\left(\frac{H}{M_s}\right)^4 \left( \frac{1}{|\xi|^5} - \frac{1}{|\xi|^3}  \right)}   
   \nonumber\\
   &\simeq  g_{\rm D3}^2\sqrt{1- 7.6\times 10^{-6} \frac{e^{2\pi \xi}}{|\xi|^3}\left(\frac{H}{M_s}\right)^4} ,   
\end{align}
where we use Eq. (\ref{eq:intensity}) with $\xi \geq {\cal O}(1)$
in the absence of the charged fermion. 
Since 
the effective coupling constant 
$G_{{\rm D}3}$ 
is not so different from the original one $g_{{\rm D}3}$
for the parameter region 
$\xi \lesssim 5$ if the mass hierarchy 
$H < M_s$ is maintained,
our discussion can be 
valid if the higher-order corrections are taken into account.

\section{Constraints on electromagnetic fields from D-brane models with axion inflation}
\label{sec:weak}

In this section, 
we show that 
inflationary models obtained from string theory satisfy the constraints 
on the electromagnetic fields.
If we consider stringy inflationary models,
several parameters can be related to the quantities in string theory
or constrained by 
the mass hierarchy.
In particular, 
the coupling constant $g$ of the electromagnetic fields, 
the axion decay constant $f $, 
the coefficient of the axion-photon coupling 
$L$,
and 
the 
cut-off scale of 
effective field theories,
will be determined or constrained by string theory.
In the following discussion, 
we show that 
there are model-independent constraints on the produced magnetic fields,
and confirm the constraints 
using some models.

\subsection{Universal expression of the instability parameter}
\label{subsec:independent}

First, we start with a model-independent analysis of estimating a typical value of 
the instability parameter $\xi$ in string axion inflation. 
By introducing the so-called slow-roll parameter:
\begin{align}
    \epsilon &= - \frac{\dot{H}}{H^2} 
    = \frac{\dot{\phi}^2}{2H^2 M_{\text{Pl}}^2},
\end{align}
the axion velocity is found to be
\begin{align}
    \dot{\phi}^2 = 2\epsilon H^2 M_{\rm Pl}^2. 
\end{align}
Therefore, 
the instability parameter $\xi$ can be expressed as 
\begin{equation}
        |\xi| 
    = \frac{L\alpha |\dot{\phi}|}{2 \pi f H} 
   = L \cdot \frac{ M_{\rm Pl} }{\sqrt{2} \pi } 
    \cdot \frac{ \alpha}{f} \cdot \sqrt{\epsilon}.
    \label{xi-r}
\end{equation}
Since the electromagnetic fields are amplified toward the end of the inflation, i.e., $\epsilon\sim {\cal O}(1)$, 
the instability parameter $\xi$ will be determined by the integer $L$ and the ratio $\fr{\alpha}{f} $.
We will show that the ratio 
$\fr{\alpha}{f} $ will be universally determined 
up to an ${\cal O}(1)$ factor, and 
$L$ is bounded from above by ${\cal O} (1)$ in typical toroidal orientifolds. 
 
First, we find that the 
ratio $\fr{\alpha}{f}$ at the tree-level
is universally determined
if the axion-photon couplings
are obtained from the Chern-Simons terms of the D-brane effective theories as previously shown in Ref.~\cite{Svrcek:2006yi}.
Here, 
the origin of the photon and the axion can be the 
gauge fields on the D-branes
and the RR
$q$-from gauge fields $C_q$
in the 10D bulk
as 
\begin{equation}
    \tilde{\phi} = \fr{2 \pi}{l_s^q}\int_{\Sigma_{q}} C_q,
\end{equation}
with a $q$-cycle $\Sigma_q$ in the extra dimensions wrapped by the D-branes, 
respectively. 
Here, the axion field $\tilde{\phi}$ is dimensionless.
Note that the rank $q$ cannot be greater 
than the dimension of D$p$-branes along the extra dimensions $q \leq p-3$ so that the D-branes can wrap $\Sigma_q$.
We will use the inequality later.

When we construct inflationary models using D-branes, 
$\alpha$ and $f$ depend on the volume moduli of extra dimensions on which the D-branes wrap. 
The volume dependence 
can be found by the kinetic terms 
of the axion and photon in 4D effective theories,
\begin{equation}
    \int d^4 x 
    \left\{\sqrt{-\det(g_{\mu \nu})}
    \left[ - 
    \fr{f^2_{4{\rm D}}}{ 2}\partial_\mu \tilde{\phi} \partial^\mu \tilde{\phi} 
    - \fr{V_{{\rm D}p}}{4 g^2_{{\rm D}p}} F_{\mu\nu}F^{\mu \nu} \right]  + \frac{L_{{\rm D}p} }{16\pi^2} 
   \tilde{\phi}
   F_{\mu \nu} \tilde{F}^{\mu \nu} + \cdots \right\}  .
\end{equation}    
Here, $f_{4{\rm D}}$ denotes the axion decay constant derived from the RR gauge fields.
$V_{{\rm D}p}$ is the volume of 
the extra dimensions on which 
the D$p$-branes wrap,
and the integer $L_{{\rm D}p}$ 
is obtained by the magnetic fluxes 
in the extra dimensions.
The dimensionless axion field 
$\tilde{\phi}$ and the axion field
$\phi $ in the previous discussions 
are related as 
$f_{\rm 4D}\tilde{\phi} = \phi$.
We have omitted other irrelevant terms.
In the following, 
we will specify these parameters.

The kinetic term of the 
axion with the decay constant 
$f_{4{\rm D}}$
can be obtained from the 
kinetic term of RR $q$-form 
gauge field in 10D bulk~(see e.g., Ref.~\cite{Blumenhagen:2013fgp}),
\begin{equation}
    S_{{\rm RR},q } = 
    - \fr{2 \pi }{l_s^{8}g_s^{(4-q )/2} }\int d^{10} x \sqrt{-g^{\rm 10D}} \fr{1}{2 (q+1)!}
    F_{A_1 ...A_{q +1} } F^{A_1 ...A_{q +1} },
\end{equation}
with the field strength 
$\fr{1}{(q+1)! }F_{A_1 ... A_{q+1}}
dx^{A_1 } \wedge \cdots \wedge dx^{A_{q+1}}
 = dC_q$,
while the kinetic term of the 
photon can be originated 
from the DBI action
on ${\rm D}p$-branes in \eqref{eq:DBI}.
The 4D axion-photon coupling can be derived from topological couplings to the RR gauge fields
on the D$p$-branes:
\begin{align}
    S_{\rm CS} = T_p \int \sum_q C_q \wedge e^{2\pi \alpha^\prime {\cal F}} \wedge \hat{A}(R),
\end{align}
where $C_q$ is pulled back onto the brane world-volume, and 
$\hat{A}(R)$ is the so-called A-roof polynomial depending on the curvature 2-form. 
Through Kaluza-Klein reduction, the 4D Chern-Simons coupling depends on the D$p$-brane worldvolume fluxes in some cases whose typical values are of ${\cal O}(1)$ denoted by $L_{{\rm D}p}$, as will be discussed below.

When the volume of $(p-3)$-cycle $V_{{\rm D}p}$ on which ${\rm D}p$-branes wrap is isotropic, i.e., $V_{{\rm D}p}= R^{p-3}$, the 4D gauge coupling constant is described by
\begin{align}
    g_{4{\rm D}}^2 = \fr{g_{{\rm D}p}^2}{V_{{\rm D}p}} = 2\pi g_s \left(\fr{l_s}{R}\right)^{p-3}.
\end{align}
The decay constant of the axion, in particular zero-modes of the $q$-form RR field $C_q$, can be read off from its kinetic term \cite{Svrcek:2006yi}
\begin{align}
    \fr{f_{4 {\rm D}}}{M_{\rm Pl}} =
    \fr{g_s}{\sqrt{8\pi^2}} \left(\fr{l_s}{R}\right)^{q},
\end{align}
Thus, the ratio 
$\alpha_{4{\rm D}} = g_{4{\rm D}}^2 / (4\pi)$ to $f_{4{\rm D}}$
can be determined as
\begin{align}
\fr{\alpha_{4{\rm D}} }{f_{4 {\rm D}}} =
\fr{\sqrt{2}\pi }{M_{\rm Pl}}
\left(\fr{l_s}{R}\right)^{p -3 -q}.
\end{align}
Since $ q \leq p - 3 $,
 the ratio must be 
$\fr{\alpha_{4{\rm D}} }{f_{4 {\rm D}}} \leq \fr{\sqrt{2}\pi }{M_{\rm Pl}}$ as long as the hierarchy $l_s < R$, i.e., 
$M_{\rm comp.} < M_s$
in \eqref{eq:hierarchy}
holds.
Since we are considering the upper bound on the instability parameter, 
we focus on the case $ p - 3  = q$,
where the ratio is universally determined as
\begin{align}
\fr{\alpha_{4{\rm D}} }{f_{4 {\rm D}}}  =  \fr{\sqrt{2}\pi }{M_{\rm Pl}}.
\label{eq:alphaf}
\end{align}
By substituting the ratio to Eq.~\eqref{xi-r}, we obtain
\begin{equation}
    |\xi | = L_{{\rm D}p} \sqrt{\epsilon}.
\end{equation}
Note that even in an anisotropic case, one can obtain the similar bound, as discussed in Ref. \cite{Svrcek:2006yi}. 
Here, we have focused on type II string theory, but the ratio $\alpha$ to $f$ is the same order in heterotic string theory. Thus, if we realize the axion inflation in heterotic string theory, as discussed in e.g., Refs. \cite{Abe:2014pwa,Ali:2014mra}, one can derive the same expression for $\xi$ in the same manner as in type II string theory.

Before going into the detail about the constraint on $L_{{\rm D}p}$, let us comment on the ratio $\fr{\alpha}{f}$. 
We have focused on the ratio $\frac{\alpha}{f}$ for the axion which appears in the gauge kinetic function at the tree-level. However, gauge kinetic functions receive moduli-dependent threshold corrections~\cite{Dixon:1990pc}. 
For instance, it was known that in type IIB string on toroidal orientifolds, axions associated with the complex structure moduli couple to the photon at the one-loop level~\cite{Blumenhagen:2006ci}.\footnote{In the case of type IIA string, see, Ref.~\cite{Lust:2003ky}.} In that case, the value of gauge coupling will be determined by the vacuum expectation value of K\"ahler moduli, and the volume dependence of $\alpha$ and $f$ for the axion of the complex structure moduli is different from each other. It results in the non-universal ratio $\frac{\alpha}{f}$. 
For the axion decay constant for the complex structure moduli, it was studied in a broad class of complex structure moduli space e.g., Refs. \cite{Conlon:2016aea,Honda:2016jnd}, which admit the small value of axion decay constant in a certain moduli space. Thus, the instability parameter $\xi$ will be enhanced by the small axion decay constant for the axion associated with the complex structure moduli.

\subsection{Bound on the instability parameter in the D-brane system without charged fermions}
\label{subsec:dependent_wo}

In this subsection, we discuss a maximum value of $L$ when the charged fermions under $U(1)$ are absent, as in the system of a single D-brane. 
For a quantitative description of the DBI bound, we define a critical value of the electromagnetic field strength: 
\begin{align}
    \langle \hat{F}_{\mu \nu}\hat{F}^{\mu \nu}\rangle\bigl|_{\rm cr} = \zeta \langle \hat{F}_{\mu \nu}\hat{F}^{\mu \nu}\rangle\bigl|_{\rm max},
    \label{eq:Fcritical}
\end{align}
where $\langle \hat{F}_{\mu \nu}\hat{F}^{\mu \nu}\rangle\bigl|_{\rm max}$ is given in \eqref{eq:Fbound}. 
Here, we introduce a dimensionless parameter $\zeta$, 
which varies from $10^{-2}$ to $10^{-4}$ in the following numerical analysis. 
Note that since the slow-roll parameter $\epsilon$ approaches to  ${\cal O}(1)$ value, 
the instability parameter is expressed by $L_{{\rm D}p}$, \eqref{eq:Fcritical} is rewritten in terms of the critical value of $L_{{\rm D}p}$, $L_{{\rm D}p}^{({\rm cr})}$,
\begin{align}
    \frac{e^{\pi L_{{\rm D}p}^{({\rm cr})}}}{|L_{{\rm D}p}^{({\rm cr})}|^{3/2}} = 
    324 \zeta \left(\frac{0.1}{g_s}\right)^{1/2} \left(\frac{M_s}{H} \right)^2.
\end{align}
From Table \ref{tab:Ncr}, $L_{{\rm D}p}^{({\rm cr})}$ is bounded from above by the DBI constraint, and the maximum value is of ${\cal O}(5)$. Since the string scale is smaller than the Planck scale, one cannot take a large hierarchy between $M_s$ and $H$. 
Such an ${\cal O}(1)$ value of $L_{{\rm D}p}$ is consistent with the string setup, as discussed below. 
\begin{table}[H]
    \centering
    \begin{tabular}{|c|c|c|c|c|}
    \hline
       $\zeta$ $\setminus$ $\frac{M_s}{H}$  &  $10^2$  & $10^3$ & $10^4$ \\ \hline
       $10^{-2}$  & 3 & 5 & 7  \\
       $10^{-3}$  & 3 & 4 & 8  \\       
       $10^{-4}$  & 2 & 3 & 5  \\ \hline      
    \end{tabular}
    \caption{Critical value of $L_{{\rm D}p}$ for $g_s = 0.1$. Here, we show a maximum integer for $L_{{\rm D}p}$. }
    \label{tab:Ncr}
\end{table}

Thus, the instability parameter as well as the Chern-Simons coupling are typically of ${\cal O}(1)$. The maximum value of primordial magnetic field at the end of inflation is also estimated as
\begin{align}
     \langle \hat{B}^2\rangle &\simeq 3.0\times 10^{-4} \frac{e^{2\pi \xi}}{|\xi|^5} H^4 \simeq 4.2 \times 10^6 H^4,   
\end{align}
where we take $\xi = 5$.

\subsection{Bound on the instability parameter in the D-brane system with charged fermions}
\label{subsec:dependent_w}

In this subsection, we move to the system with charged massless fermion under $U(1)$, which can be realized in magnetized D-branes.  
By using the critical value of the electromagnetic fields \eqref{eq:Fcritical}, 
\eqref{eq:massless-upper-bound} is rewritten in terms of $L_{{\rm D}p}$ for the slow-roll parameter $\epsilon={\cal O}(1)$,
\begin{equation}
 2(12\pi^2 L_{{\rm D}p}^{({\rm cr})})^2 
    =15.6 \zeta 
    \left( 
    \frac{g_s}{0.1} 
 \right)^2 \left( 
    \fr{M_s}{H} 
 \right)^4 .
\end{equation}
In contrast to the system without charged fermions, one can consider a large value of $L_{{\rm D}p}^{({\rm cr})}$ for $M_s\gg H$. 
However, a large value of $L_{{\rm D}p}^{({\rm cr})}$ induces a large amount of D-brane charges, which will be in tension with the tadpole cancellation condition of D$p$-branes. 
To clarify this statement, we discuss the constraint on $L_{{\rm D}p}$ from a different point of view.

\paragraph{Constraints on axion-photon coupling $L$}\,\\

Let us consider the $T^6/(\mathbb{Z}_2\times \mathbb{Z}_2^\prime)$ orientifold with O3/O7-planes. (For a detailed discussion, see, Refs. \cite{Blumenhagen:2006ci,Ibanez:2012zz}.) 
For $N_a$ stacks of magnetized D$7_a$-branes, wrapping numbers $n_a^i$ and magnetic fluxes $m_a^j$ on $(T^2)_i$, with $i=1,2,3$, are defined as\footnote{Here, $n_a^i$ and $m_a^i$ for each $a,i$ are chosen as coprime numbers.
This field strength should be described as
\[
F_a^i = 
\left(\begin{matrix}
\frac{2 \pi m_a^i}{n_a^i}
\boldsymbol{1}_{n_a^i }  &  
\\  
& 
\boldsymbol{0}_{N_a - n_a^i}
\end{matrix}
\right)
\]
The Dirac quantization condition for
the $U(1)$ factor is 
$\int_{(T^2)_i} {\rm tr} F_a^i = 2\pi m_a^i$.
}
\begin{align}
    \frac{n_a^i}{2\pi} \int_{(T^2)_i} F_a^i = m_a^i,
\end{align}
where $F_a$ corresponds to $U(1)_a$ gauge field strength on D$7_a$-branes. 
Such topological numbers appear in a gauge kinetic function on D$7_a$-branes wrapping $(T^2)_j\times (T^2)_k$:
\begin{align}
    2\pi f^{{\rm D}7_a} = m_a^j m_a^k S -  n_a^j n_a^k T_i, \qquad i\neq j \neq k,
\end{align}
where the axionic component of the axio-dilaton $S$ and the K\"ahler moduli $T_i$ (determining the volume of $(T^2)_j\times (T^2)_k$) originate from the Kaluza-Klein reduction of RR fields $C_0$ and $C_4$, respectively. 
Thus, a product of wrapping numbers $n^j$ and magnetic fluxes $m^j$ controls the magnitude of topological coupling $L_{{\rm D}p}$. 
However, these topological numbers are severely restricted to satisfy the cancellation of D-brane charge:
\begin{align}
    \sum_a N_a m_a^1 m_a^2 m_a^3 + \frac{N_{\rm flux}}{2} = 16,
\end{align}
for D3-brane and 
\begin{align}
    \sum_a N_a m_a^i n_a^j n_a^k = - 16,
\end{align}
for D$7_a$-branes wrapping $(T^2)_j\times (T^2)_k$. 
Here, $N_{\rm flux}$ denotes the tadpole charge determined by imaginary self-dual three-form fluxes. 
Note that the 64 O3-plane and 4 O7-planes contribute to RR charges $-32$ in corresponding homology classes. 
Since the generation number of quarks and leptons is determined by the combination of wrapping numbers and magnetic fluxes, we have to be careful about the choice of these topological numbers.

For instance, for the local supersymmetric brane configurations incorporating the Standard-Model spectra \cite{Marchesano:2004xz}\footnote{Note that the $U(1)_{B-L}$ is identified as a linear combination of $U(1)_a$ and $U(1)_d$, and the other $U(1)$ becomes massive through the Green-Schwarz mechanism. The hypercharge $U(1)_Y$ can be realized as a spontaneous symmetry breaking of $SU(2)_R\times U(1)_{B-L}$.}, the topological numbers are specified as
\begin{table}[H]
    \centering
    \begin{tabular}{|c|c|c|c|c|} \hline
        $N_\alpha$ & {\rm Gauge~group} & ($m_\alpha^1, n_\alpha^1$) & ($m_\alpha^2, n_\alpha^2$) & ($m_\alpha^3, n_\alpha^3$)\\ \hline \hline
        $N_a=6$ & $U(3)_C$ & (1,0) & ($m_{\rm flux}, 1$) & ($m_{\rm flux}, -1$)\\ \hline
         $N_b=2$ & $SU(2)_L$ & (0,1) & ($1, 0$) & ($0, -1$)\\ \hline
        $N_c=2$ & $SU(2)_R$ & (0,1) & ($0, -1$) & ($1,0$)\\ \hline
        $N_d=2$ & $U(1)_d$ & (1,0) & ($m_{\rm flux}, 1$) & ($m_{\rm flux}, -1$)\\ \hline
    \end{tabular}
\end{table}
Then, the gauge kinetic functions of D7-branes are given by
\begin{align}
    {\rm D}7_a &: 2\pi f_a = m_{\rm flux}^2 S + T_1,
    \nonumber\\
    {\rm D}7_b &: 2\pi f_b = T_2,
   \nonumber\\
    {\rm D}7_c &: 2\pi f_c = T_3,
   \nonumber\\
    {\rm D}7_d &: 2\pi f_d = m_{\rm flux}^2 S + T_1.
\label{eq:fD7}
\end{align}
It indicates that topological couplings to RR fields $L_{{\rm D}p}$ can be enhanced by the magnetic flux $m_{\rm flux}$. 
However, the possible choices of magnetic flux $m_{\rm flux}$ is of ${\cal O}(1)$ through  the tadpole cancellation condition of D$3$-branes:
\begin{align}
    m_{\rm flux}^2 = 2 - \frac{N_{\rm flux}}{16},
\end{align}
where it is notable that $N_{\rm flux}$ is positive due to the imaginary self-dual condition. 
Thus, the magnitude of topological couplings to RR coupling $L_{{\rm D}p}$ is of ${\cal O}(1)$.

So far, we have focused on a specific toroidal orientifold compactification with Standard-Model spectra, but it can be generalized to D$p$-branes wrapping holomorphic divisors in Calabi-Yau manifolds. 
As an extension of toroidal orientifolds, let us consider generic type IIB orientifolds with O3/O7-planes, which can be uplifted to F-theory on Calabi-Yau fourfolds. 
It was known that the D3-brane charges should be canceled to satisfy the so-called tadpole cancellation condition \cite{Becker:1996gj,Sethi:1996es,Haack:2001jz}:
\begin{align}
    N_{\rm D3} + \frac{N_{\rm flux}}{2} + N_{{\rm D}7}^{\rm flux} = \frac{\chi (Y_4)}{24},
\end{align}
where $N_{\rm D3}$, $N_{\rm flux}$, $N_{{\rm D}7}^{\rm flux}$ and $\chi (Y_4)$ denote the number of D3-branes, D3-brane charge induced by three-form fluxes and magnetic fluxes on D7-branes, and Euler characteristics of Calabi-Yau fourfold, respectively.

When we consider $U(1)$ magnetic fluxes on $N$ stacks of D7-branes, the gauge flux can be expanded in the basis of D7-brane divisors $D_a$ with $a=1,2,...,h^{1,1}(D_a)$. 
Let us denote $c_1(l_a)$ as the first Chern class of line bundle $l_a$ with structure group $U(1)$, which is quantized as $c_1(l_a) = m_a [D_a]$ on each divisor. Note that $m_a$ determines the magnitude of the axion-photon coupling, which appears in the gauge kinetic function of D7-branes, as in \eqref{eq:fD7}. 
Then, the contribution of gauge flux $ N_{{\rm D}7}^{\rm flux}$ is rewritten as \cite{Blumenhagen:2008zz}
\begin{align}
    N_{{\rm D}7}^{\rm flux} &= \frac{1}{2}\sum_a N_a m_a^2 \kappa_{aaa},
\end{align}
where $N_a$ is the number of D7-branes wrapping holomorphic divisor $D_a$, and $\kappa_{aaa}$ is defined as 
$\kappa_{aaa}= -\int [D_a]\wedge [D_a]\wedge [D_a]$. 
As discussed in Ref.~\cite{Ishiguro:2023jjc}, the $ N_{{\rm D}7}^{\rm flux}$ is bounded from above by using the 
Cauchy-Schwarz inequality:
\begin{align}
    |m_{\rm max}|^2 \leq \frac{\chi (Y_4)}{12\,{\rm min}_a(N_a \kappa_{aaa})} - (h^{1,1}(D_a)-1), 
\end{align}
where we define $m_{\rm max}$ such that $|m_a| \leq {\rm max}_{a}(m_a):=|m_{\rm max}|$ for all $a$, 
and we assume $m_a^i\neq 0$ for all $a$.

In a similar to the toroidal case, the magnitude of topological couplings to RR fields is mutually related with the tadpole cancellation condition~\cite{Ishiguro:2023jjc}. 
In the context of IIB/F-theory, the possible value of flux quanta depends on the Euler characteristics of Calabi-Yau fourfolds. 
Since the largest value of $\chi (Y_4)$ is known as 1,820,448~\cite{Candelas:1997eh,Taylor:2015xtz}, it seems to consider the large 
value of flux quanta, in other words the large value of axion-photon coupling $L_{{\rm D}p}$, in such a Calabi-Yau compactification. 
However, there exists the large number of moduli fields on Calabi-Yau fourfolds with large value of $\chi (Y_4)$. 
Of course, one can consider the large value of background fluxes in Calabi-Yau manifolds with large number of moduli, but it would be difficult to stabilize all the moduli fields at the stable vacua, as in the tadpole conjecture \cite{Bena:2020xrh}.
Furthermore, from the phenomenological viewpoint, the value of magnetic fluxes is related to the number of chiral zero-modes on D7-branes through the index theorem. 
It indicates that the small value of magnetic fluxes is also favored to be consistent with the low-energy effective theory including the Standard Model. 
We conclude that it is in general difficult to realize 
the electromagnetic fields that are much greater than 
the Hubble scale, which is consistent with 
the mass hierarchy involving the volume of the extra dimensions and the string scale. 
The instability parameter as well as the Chern-Simons coupling are typically of ${\cal O}(1)$. The maximum value of primordial magnetic field at the end of inflation is estimated by using \eqref{eq:Emax_wfermion}
\begin{align}
     \langle \hat{B}_{\rm max}\rangle &\simeq 1.2\times 10^5 \left(\frac{10^{-1}}{g}\right)^3 H^2.
\end{align}

\section{Conclusions}
\label{sec:con}

A tachyonic production of gauge field helicities during the axion inflation 
has many attractive features such as the production of density perturbations and gravitational waves through 
the topological coupling between the electromagnetic fields and the axion. 
The magnitude of one gauge field helicity during the axion inflation can be controlled by the parameter $\xi$, which can 
be expected to be ${\cal O}(1)$ to probe the axion dynamics by the cosmological observations. 

In this paper, we examined stringy constraints for produced electromagnetic fields induced by the non-zero axion velocity during 
the inflation. 
First, we discuss the validity of the model with conventional quadratic kinetic terms of axion and photon. 
Since the action of axion electrodynamics will be regarded as the low-energy effective description of DBI action in string theory, 
the intensity of electromagnetic fields is bounded from above by the string scale to have control over the effective field theory. 
It gives the constraint on $\xi$ as shown in Eq. (\ref{eq:xi}), whose bound can be rewritten in terms of supersymmetry 
breaking scale in the context of type IIB Large Volume Scenario. 
Second, we take into account the tadpole constraints.
Given that the ratio the gauge coupling to the axion decay constant is constant in weakly-coupled string theories, 
we found that the parameter $\xi$ was given by the slow-roll parameter $\epsilon$: $|\xi| \simeq L_{{\rm D}p} \sqrt{\epsilon}$ where we use the slow-roll approximation in the axion inflation. 
In the analysis of Sec. \ref{subsec:dependent_wo} where the charged fermions under $U(1)$ absent in the system, the coefficient of axion-photon coupling $L_{{\rm D}p}$ is restricted to ${\cal O}(1)$ due to the hierarchy between the string scale and the inflation scale. 
They indicate that it will be in general difficult to realize ${\cal O}(1)$ value of $\xi$ in the string effective theory. 
On the other hand, the existence of charged fermions allows us a large value of the instability parameter, but it is constrained to satisfy the tadpole cancellation condition as well as the tadpole conjecture \cite{Bena:2020xrh}, as discussed in Sec. \ref{subsec:dependent_w}. 
Hence, we conclude that it is in general difficult to realize a sizable amount of primordial electromagnetic fields in stringy axion inflation.

\acknowledgments

We thank K. Kamada and K. Mukaida for helpful discussions and comments. 
This work was supported by JSPS KAKENHI Grant Numbers JP20K14477 (H.~O.), JP21J00480 (R.~Y.), JP21K13928 (R.~Y.), JP23H04512 (H.~O).

\bibliography{referencesv2}{}
\bibliographystyle{JHEP}

\end{document}